# Localization of transverse waves in randomly layered media at oblique incidence

## K.Yu. Bliokh[1,2*] and V.D. Freilikher[2,3]

[1]*Institute of Radio Astronomy, 4 Krasnoznamyonnaya st., Kharkov, 61002, Ukraine*
[2]*Department of Physics, Bar-Ilan University, Ramat-Gan, 52900, Israel*
[3]*Complex Photonic Systems, Department of Science and Technology and MESA+ research institute, University of Twente, P.O. Box 217, 7500 AE Enschede, The Netherlands*

We investigate the oblique incidence of electromagnetic waves on a randomly layered medium in the limit of strong disorder. An approximate method for calculating the inverse localization length based on the assumptions of zero energy flux and complete phase stochastization is presented. Two effects not found at normal incidence have been studied: dependence of the localization length on the polarization, and decrease of the localization length due to the internal reflections from layers with small refractive indexes. The inverse localization length (attenuation rate) for $P$-polarized radiation is shown to be always smaller than that of $S$-waves, which is to say that long enough randomly layered sample polarizes transmitted radiation. The localization length for $P$-polarization depends non-monotonically on the angle of propagation, and under certain conditions turns to infinity at some angle, which means that typical (non-resonant) random realizations become transparent at this angle of incidence (stochastic Brewster effect).

## 1. INTRODUCTION

Localization of classical waves and quantum particles in one-dimensional (1D) disordered systems is well studied theoretically (see, for example, [1], [2] and references therein). Important application of the theory is the propagation of radiation in randomly layered media where the refractive index depends on a single coordinate. In general, however, this is a three-dimensional problem, which is reduced to a 1D one only when the direction of propagation is normal to the layers. In this case the field, $\psi(z)$, of a monochromatic wave obeys a Schrödinger-type equation with energy-dependent effective potential

$$-\frac{d^2\psi(z)}{dz^2} + k^2 \delta\varepsilon(z)\psi(z) = k^2 \psi(z) \ , \tag{i}$$

where $z$ axis is normal to the layers, $k = \sqrt{\varepsilon_0}\,\omega/c$, $\omega$ is the frequency, $\delta\varepsilon(z) = -\Delta\varepsilon(z)/\varepsilon_0$, and the dielectric constant of the medium is given by

$$\varepsilon(z) = \varepsilon_0 + \Delta\varepsilon(z) \ , \tag{ii}$$

with $\Delta\varepsilon(z)$ being a random function of the coordinate. The important distinction of Eq. (i) from the corresponding quantum-mechanical equation for electrons is that the "energy", $k^2$, in Eq. (i) is always higher than the "potential", $k^2\delta\varepsilon$, (unless $\Delta\varepsilon/\varepsilon_0 < -1$), i.e. only "above-barrier" scattering takes place. In other words, total internal reflection never occurs at normal incidence, and localization in this case is due to the interference of the multiply scattered random fields.

---
[*] E-mail: kostya@bliokh.kharkiv.com, k_bliokh@mail.ru

The situation is, however, different when oblique (with respect to $z$-axis) incidence is considered. In this instance the field can be presented as $\psi(\mathbf{r}) = \exp(ik_x x + ik_y y)\chi(z)$, ($k_x$ and $k_y$ are $x$- and $y$- components of the wave vector respectively), and the equation for the $z$-dependent term, $\chi(z)$, takes the form

$$-\frac{d^2\chi(z)}{dz^2} + k^2\delta\varepsilon(z)\chi(z) = \left(k^2 - k_x^2 - k_y^2\right)\chi(z) \quad . \tag{iii}$$

Obviously the "energy", $k^2 - k_x^2 - k_y^2$, may take any positive value, in particular can be less than "potential" $k^2\delta\varepsilon$. It gives rise to an additional mechanism of localization, which is due to the internal reflection and tunnelling.

Another new effect, which is absent in pure 1D random systems, comes about at oblique propagation of transverse vector waves. In this case the symmetry with respect to the direction of propagation is broken, and the localization length depends significantly on the polarization of the radiation. This phenomenon is a consequence of the dependence of Fresnel reflection and transmission coefficients on the wave polarization [3].

In this paper we present an approximate method for calculating the localization length in randomly layered medium based on the assumptions of the exponentially small transmission and complete phase randomization (Sec. 2). In Sec. 3 we use this method to calculate the localization length for two orthogonal linear polarizations. It is shown that the localization length of the wave with the vector of the electric field orthogonal to the plane of incidence ($S$ -wave) is always larger than that of $P$ -waves (with the electric vector in the plane of incidence), for which a sort of stochastic Brewster effect takes place. As the result, the radiation transmitted through a long enough randomly layered sample is always $P$ -polarized (with an exponential accuracy). The effect on the localization length of the internal reflection at the interfaces between random layers is studied in Sec. 4. Some examples of randomly layered media have been considered.

## 2. METHOD OF CALCULATION OF THE LOCALIZATION LENGTH IN A RANDOMLY LAYERED MEDIUM

It is well known that the modulus of the reflection coefficient, $R$, of a plane monochromatic wave incident on a randomly layered half-space is equal to one, and there is no energy flux inside the medium generated by the incident wave [1,2,4]. When a randomly layered sample has a finite but large enough length, $L$, then $|R(L)|$ differs from unity by an exponentially small number, $1 - |R(L)|^2 \propto \exp(-2L/l_{loc})$, ($L \gg l_{loc}$, $l_{loc}$ is the localization length), and the flux along the system is exponentially small, $\propto \exp(-2L/l_{loc})$. This *a priori* information enables to assume that if a plane wave with frequency $\omega$ is incident normally (along $z$ axis) on a sample from left, the field in each $j$ th layer inside the sample can be considered (with an exponential accuracy) a standing wave, and presented as

$$\psi_j = A_j \exp(i\omega t)\cos\left[k_j\left(z - z_j\right) + \varphi_j\right] \quad . \tag{1}$$

Here $A_j$ is the real amplitude, $k_j = n_j\omega/c$, $n_j$ is the refractive index, $\varphi_j$ is the (real) phase at the right-hand boundary of the layer located at a point $z_j$. Such presentation of the phase is dictated by the fact that the transmission problem for the wave incident from left can be formulated as an evolutional one with initial conditions given at the right edge of the sample [4,5].

Thus the wave propagation problem is reduced to the oscillatory one with single unknown real amplitude and real phase, Eq. (1). This simplifies the problem significantly as compared to the conventional transfer matrix method [6−9], where the evolution of two independent waves in each layer is considered. To calculate the localization length we use the standard definition [1]

$$l_{loc}^{-1} = -\left\langle \frac{\ln T}{2L} \right\rangle = -\lim_{L \to \infty} \left( \frac{\ln T}{2L} \right) , \qquad (2)$$

where $T$ is the transmission coefficient of a random sample. Notice that the inverse localization length, $l_{loc}^{-1}$, is a self-averaging quantity, which means that the value measured at any finite but long enough realization coincides with the exponential accuracy with its mean value [1].

We represent $\ln T$ as

$$\ln T = \ln \left( \frac{A_{N+1}}{A_0} \right)^2 = 2 \ln \left( \frac{A_{N+1}}{A_N} \frac{A_N}{A_{N-1}} ... \frac{A_1}{A_0} \right) = 2 \sum_{j=0}^{N} \ln \frac{A_{j+1}}{A_j} , \qquad (3)$$

where $N$ is the total number of layers, $A_0$ and $A_{N+1}$ are field amplitudes to the left and to the right of the sample correspondingly. By substituting Eq. (3) into Eq. (2) we obtain

$$l_{loc}^{-1} = -\lim_{N \to \infty} \left( L^{-1} \sum_{j=0}^{N} \ln \frac{A_{j+1}}{A_j} \right) . \qquad (4)$$

Note that $\lim_{N \to \infty} L / N = \langle s_j \rangle \equiv s$, where $s_j$ is the thickness of $j$th layer, and $s$ is the mean thickness of the layers. Since $\lim_{N \to \infty} \left[ N^{-1} \sum_{j=0}^{N} \ln \left( A_{j+1} / A_j \right) \right] = \langle \ln \left( A_{j+1} / A_j \right) \rangle$, Eq. (4) becomes

$$l_{loc}^{-1} = s^{-1} \left\langle \ln \frac{A_j}{A_{j+1}} \right\rangle . \qquad (5)$$

Thus, the inverse localization length $l_{loc}^{-1}$ is approximately equal to the inverse average thickness of layers times the mean logarithm of the ratio of field amplitudes in adjacent layers. The connection between $(A_j, \varphi_j)$ and $(A_{j+1}, \varphi_{j+1})$ should be found from the boundary conditions at the corresponding interfaces, and can be written in the most general form as

$$A_j = A_{j+1} f \left( \varphi_{j+1} - \Delta_{j+1}, \mu_j, \mu_{j+1} \right) , \quad \varphi_j = g \left( \varphi_{j+1} - \Delta_{j+1}, \mu_j, \mu_{j+1} \right) . \qquad (6)$$

Here $\Delta_j = k_j s_j$ is the phase increment in the $j$th layer, $f$ and $g$ are some functions, and $\mu_j$ is a set of parameters, describing the random system (refractive index, impedance, density, etc.). The first Eq. (6) shows the linear connection of the amplitudes due to linearity of the problem. The second one accounts for the phase change at the interface between two neighboring layers (see Eq. (1) and the paragraph below). From Eqs. (6), (5) it follows that

$$l_{loc}^{-1} = s^{-1} \left\langle \ln f \left( \varphi_{j+1} - \Delta_j; \mu_j, \mu_{j+1} \right) \right\rangle , \qquad (7)$$

where $<...>$ stands for the averaging over all random parameters ($s_j$ and $\mu_j$) the function $f$ depends on. The explicit form of the distribution of phases $\varphi_j$ should be found from boundary conditions (6), which is rather formidable task in general case. The problem, however, is simplified significantly in the high-frequency limit, when the disorder is strong enough that the phases ($\varphi_j - \Delta_j$) can be considered as independent random variables homogeneously distributed in the interval $(0, 2\pi)$ [1,4,6−11]. In this instance Eq. (7) takes the form

$$l_{loc}^{-1} = s^{-1} \iint d\mu' d\mu'' P(\mu') P(\mu'') \frac{1}{2\pi} \int_0^{2\pi} d\varphi' \ln f \left( \varphi', \mu'', \mu' \right) , \qquad (8)$$

where $P(\mu) d\mu$ is the joint probability density distribution of parameters $\mu$. Eq. (8) enables calculation of the inverse localization constant in the high-frequency limit for linear waves of any nature (electromagnetic, acoustic, or seismic, etc.). The method is rather general, and can be easily modified for different kinds of random systems, for instance, periodical in some parameter(s), or those containing several types of layers with distinct dielectric properties and statistics. Examples of such systems are considered in Sec. 3 and Sec. 4.

## 3. POLARIZATION DEPENDENCE

In this section we study the oblique propagation of electromagnetic waves in a passive dielectric medium, which consists of alternating sequence of layers with random thicknesses, $s_i$, and random real dielectric constants, $\varepsilon_j$. The electric, **E,** and magnetic, **H**, fields of the wave are described by Maxwell equations

$$\operatorname{curl}\mathbf{E} - ik_0\mathbf{H} = 0 ,$$
$$\operatorname{curl}\mathbf{H} + ik_0\varepsilon(z)\mathbf{E} = 0 , \tag{9}$$

where $k_0 \equiv \omega/c$ and $\varepsilon(z)$ is the generalized telegraphic random process. Without loss of generality we assume that $k_y = 0$, and consider two mutually orthogonal waves with $H_y = E_x = E_z \equiv 0$, $E_y = E(z)$ and $E_y = H_x = H_z = 0$, $H_y = H(z)$ ($S$- and $P$- waves respectively). These waves are the independent eigen modes of the one-dimensionally inhomogeneous medium that propagate without interaction and change of polarization. After substitution

$$\mathbf{E} \to \mathbf{E}(z)\exp(ik_x x) ,$$
$$\mathbf{H} \to \mathbf{H}(z)\exp(ik_x x) , \tag{10}$$

the system of equations (9) can be reduced to two independent equations for non-zero components of $P$- and $S$- waves:

$$\frac{d^2 E}{dz^2} + k_z^2 E = 0 , \tag{11}$$

$$\frac{d^2 H}{dz^2} - \frac{1}{\varepsilon}\frac{d\varepsilon}{dz}\frac{dH}{dz} + k_z^2 H = 0 . \tag{12}$$

Here

$$k_z(z) = \sqrt{k_0^2 n^2(z) - k_x^2} = k_0\sqrt{n^2(z) - n_c^2} , \tag{13}$$

where $n(z) = \sqrt{\varepsilon(z)}$ is the (random) refractive index and $n_c = k_x/k_0 = \sin\vartheta_0$ is its critical value ($\vartheta_0$ is the angle between the wave vector and $z$ axis in vacuum). A layer is transparent when $n > n_c$ and reflecting when $n < n_c$. To calculate the inverse localization length for $S$-waves, $l_{loc_S}^{-1}$, we use the general result of Sec. 2, namely, Eq. (8), where the explicit expression for the function $f$ should be substituted. In accordance to Eq. (1), the electric field $E$ of the $S$-wave in $j$th layer (solution of Eq. (11)) can be written as

$$E_j = A_j \cos\left[k_{z_j}(z - z_j) + \varphi_j\right] , \tag{14}$$

Then, from the continuity conditions for the tangential component of the electric field and its normal derivative $dE_y/dz \sim H_x$ at the interface between $j$th and $j+1$th layers it follows:

$$A_j \cos\varphi_j = A_{j+1}\cos\left(\varphi_{j+1} - \Delta_{j+1}\right),$$
$$A_j k_{z_j}\sin\varphi_j = A_{j+1}k_{z_{j+1}}\sin\left(\varphi_{j+1} - \Delta_{j+1}\right). \tag{15}$$

From Eq. (15) we derive

$$A_j = A_{j+1} f_S \left( \varphi_{j+1} - \Delta_{j+1}, n_j, n_{j+1} \right), \quad f_S = \left[ \cos^2 \left( \varphi_{j+1} - \Delta_{j+1} \right) + \frac{k_{z_{j+1}}^2}{k_{z_j}^2} \sin^2 \left( \varphi_{j+1} - \Delta_{j+1} \right) \right]^{\frac{1}{2}}, \quad (16)$$

(see Eq. (6)). Phase averaging, i.e. the integration over variable $\varphi$ in Eq. (8) with $f$ given by Eq. (16), can be performed taking into account that

$$\frac{1}{2\pi} \int_0^{2\pi} \ln \left( a^2 \sin^2 \varphi + b^2 \cos^2 \varphi \right) d\varphi = \ln \frac{(a+b)^2}{4} \ .$$

This gives

$$l_{loc_S}^{-1} = s^{-1} \int_{n' > n_c} dn' F_n(n') \int_{n'' > n_c} dn'' F_n(n'') \ln \left\{ \frac{1}{2} \left[ 1 + \frac{k_z(n')}{k_z(n'')} \right] \right\} \ . \quad (17)$$

Here $k_z(n)$ is given by Eq. (13), $F_n$ is the distribution function of the refractive indices, $n_j$, of the layers. The limits of integration in Eq. (17) follow from the simplifying assumption that all layers are transparent. The effect of reflecting layers is considered in Sec. 4.

Inverse localization length for $P$-waves, $l_{loc_P}^{-1}$ can be calculated in the same way, by using the continuity conditions for the tangential component of the magnetic field $H$ (Eq. (12)) and for $\varepsilon^{-1} dH_y / dz \sim E_x$. It is easy to show that in this instance

$$A_j = A_{j+1} f_P \left( \varphi_{j+1} - \Delta_{j+1}, n_j, n_{j+1} \right), \quad f_P = \left[ \cos^2 \left( \varphi_{j+1} - \Delta_{j+1} \right) + \frac{k_{z_{j+1}}^2 n_j}{k_{z_j}^2 n_{j+1}} \sin^2 \left( \varphi_{j+1} - \Delta_{j+1} \right) \right]^{\frac{1}{2}}. \quad (18)$$

Note that the expression for $f_P$, Eq. (18), coincides with that for $f_S$, Eq. (20), after substitution $k_{z_j} \to k_{z_j} / n_j^2$. Therefore, the same substitution in Eq. (17) yields

$$l_{loc_P}^{-1} = s^{-1} \int_{n' > n_c} dn' F_n(n') \int_{n'' > n_c} dn'' F_n(n'') \ln \left\{ \frac{1}{2} \left[ 1 + \frac{n''^2 k_z(n')}{n'^2 k_z(n'')} \right] \right\} \ . \quad (19)$$

Eqs. (17), (19) show that the localization lengths of $S$- and $P$- waves are different. To compare them it is convenient to rewrite expressions (17), (19) in the symmetrical with respect to the integration variables $n'$ and $n''$ form:

$$l_{loc_S}^{-1} = (2s)^{-1} \int_{n' > n_c} dn' F_n(n') \int_{n'' > n_c} dn'' F_n(n'') \ln A(n', n''), \quad A = \frac{[k_z(n') + k_z(n'')]^2}{4 k_z(n') k_z(n'')} \ , \quad (20)$$

$$l_{loc_P}^{-1} = (2s)^{-1} \int_{n' > n_c} dn' F_n(n') \int_{n'' > n_c} dn'' F_n(n'') \ln B(n', n''), \quad B = \frac{[n''^2 k_z(n') + n'^2 k_z(n'')]^2}{4 n'^2 n''^2 k_z(n') k_z(n'')} \ . \quad (21)$$

Easy to show that

$$d = A - B = \frac{\left( n'^2 - n''^2 \right)^2 \sin^2 \vartheta_0}{4 n'^2 n''^2 k_z(n') k_z(n'')} \geq 0 \ , \quad (22)$$

which means that

$$l_{loc_S}^{-1} \geq l_{loc_P}^{-1} \ , \quad (23)$$

The equality in Eq. (23) corresponds to the "degenerate" cases of normal incidence, $\vartheta_0 = 0$, or to a homogeneous medium, $F_n(\zeta) = \delta(\zeta - n)$.

Thus, the $S$-wave is always stronger localized than the $P$-wave. From Eqs. (20), (21) it also follows that the transmission coefficients are different for different polarizations $T_{S,P} = \exp \left( -2L / l_{loc_{S,P}} \right)$. It means that randomly layered medium acts as a *polarizer* for an obliquely propagating radiation. Indeed, if an incident wave has a mixed polarization, but the thickness of the randomly layered slab is large enough, the transmitted wave will be (with exponential accuracy) $P$-polarized.

The difference in localization lengths grows with the increase of the angle of incidence. For example, at small angles of propagation, $\vartheta_0 \ll 1$, $\sin \vartheta_0 = n_c \ll n_j$, it can be readily shown from Eqs. (20), (21) that

$$l_{loc_{S,P}}^{-1} = l_{loc_0}^{-1} \pm a\,\vartheta_0^2 + O\left(\vartheta_0^4\right)\,, \quad a = (2s)^{-1}\int dn' F_n(n')\int dn'' F_n(n'')\,\frac{(n'-n'')^2}{2n'^2 n''^2}\,, \tag{24}$$

where $l_{loc_0}^{-1}$ is the inverse localization length at normal propagation ($k_x = \sin \vartheta_0 = 0$, $k_z = k = k_0 n$). One can see that when $\vartheta_0$ increases, $l_{loc_S}^{-1}$ grows and $l_{loc_P}^{-1}$ decreases proportionally to $\vartheta_0^2$ and symmetrically with respect to $l_{loc_0}^{-1}$. Obviously, the following inequality holds

$$l_{loc_S}^{-1} > l_{loc_0}^{-1} > l_{loc_P}^{-1}\,. \tag{25}$$

Rise of the localization length (weakening of localization) of $P$-wave with $\vartheta_0$ increasing stems from the decrease of the reflection coefficient from the interface between two homogeneous media [3]. If the media are infinite the reflection coefficient for $P$-wave becomes zero at $\vartheta_0 = \vartheta_B$, where $\vartheta_B$ is so-called Brewster angle. In general, in the case of randomly layered medium the reflection coefficient does not turn to zero, however, the inverse localization length $l_{loc_P}^{-1}$ reaches a minimum at some angle $\vartheta_0 = \tilde{\vartheta}_B$ that can be found from the condition

$$\left.\frac{dl_{loc_P}^{-1}}{d\vartheta_0}\right|_{\vartheta_0 = \tilde{\vartheta}_B} = 0\,. \tag{26}$$

The dependences $l_{loc_{S,P}}^{-1}\left(\vartheta_0\right)$ for the case of rectangle distribution function $F(n)$ (see Sec. 4) is shown in Fig. 1.

Interestingly, in the particular case of a layered medium built of alternating layers of two dielectrics with refractive indices $n_1$ and $n_2$ and random thicknesses, there exist Brewster angle, $\vartheta_0 = \vartheta_B$, at which the reflection coefficient of $P$-wave turns to zero and localization is absent: $l_{loc_P}^{-1} = 0$. In this case the inverse localization lengths can be calculated explicitly. To do this, boundary conditions (15), (16) should be applied twice: for the transition from a layer $n_1$ to the adjacent layer $n_2$, and from the layer $n_2$ to the next layer $n_1$. The similar problem for normal propagation was solved in [6–11]. By multiplying together two equations like Eq. (16) we find:

$$A_{j-1} = A_{j+1}\tilde{f}_S\left(\varphi_{j+1} - \Delta_{j+1}, \varphi_j - \Delta_j, n_1, n_2\right),$$

$$\tilde{f}_S = \left[\cos^2\left(\varphi_{j+1} - \Delta_{j+1}\right) + \frac{k_{z_1}^2}{k_{z_2}^2}\sin^2\left(\varphi_{j+1} - \Delta_{j+1}\right)\right]\left[\cos^2\left(\varphi_j - \Delta_j\right) + \frac{k_{z_2}^2}{k_{z_1}^2}\sin^2\left(\varphi_j - \Delta_j\right)\right], \tag{27}$$

where $k_{z_{1,2}}^2 = k_0^2 n_{1,2}^2 - k_x^2$. After substitution Eq. (27) into an equation similar to Eq. (8) and averaging over phases $\varphi_{j+1} - \Delta_{j+1}$ and $\varphi_j - \Delta_j$ we obtain

$$l_{loc_S}^{-1} = \frac{1}{s_0}\ln\frac{\left(k_{z_1} + k_{z_2}\right)^2}{4k_{z_1}k_{z_2}}\,, \tag{28}$$

$$l_{loc_P}^{-1} = \frac{1}{s_0}\ln\frac{\left(k_{z_1} n_2^2 + k_{z_2} n_1^2\right)^2}{4k_{z_1}k_{z_2}n_1^2 n_2^2}\,. \tag{29}$$

Here $s_0 = s_1 + s_2$ is the mean thickness of the pair of the layers, $s_{1,2}$ are the mean thicknesses of the layers with refractive indices $n_{1,2}$ respectively. If $k_{z_1} n_2^2 = k_{z_2} n_1^2$ the inverse localization length of $P$-wave, Eq. (29) turns to zero. This determines Brewster angle for the considering layered medium:

$$\vartheta_B = \frac{n_1 n_2}{\sqrt{n_1^2 + n_2^2}} \ . \tag{30}$$

If $\vartheta_0 = \vartheta_B$, localization is absent for $P$-wave. This circumstance is related to the fact that quantity $k_z / n^2$ plays role of the effective longitudinal wave number of $P$-wave. Then, if $k_{z_1} / n_1^2 = k_{z_2} / n_2^2$ the medium is effectively homogeneous.

## 4. EFFECT OF THE INTERNAL REFLACTION

In Sec. 2 the localization length have been calculated under the assumption that the refractive index of all random layers was larger than a critical value $n_j > n_c = k_x / k_0 = \sin \vartheta_0$. It guaranteed that the angle of incidence at any interface between $j$th and $(j+1)$th layers was always smaller than the angle of total internal reflection, $\vartheta_j^{(tot)}$, which corresponded to the above-barrier reflection of a quantum particle. In general case, however, this restriction must be removed, i.e. for some layers inside the random system the inverse inequality, $n_j < n_c$, may take place. It means that for these layers the local angle of incidence exceeds $\vartheta_j^{(tot)}$, and strong internal reflection from them should be taken into account. In what follows we call such layers 'reflecting' to distinguish them from the 'transparent' (with no internal reflection) ones. The longitudinal wave number, $k_{z_j}$, Eq. (13), inside $j$th reflecting layer is an imaginary number, therefore the wave exponentially decays along $z$-axis. Nevertheless the transmission coefficient is finite, and a propagating wave with finite amplitude is incident on $(j+1)$th layer (we assume that $n_{j+1} > n_c$). Note that even small amount of reflecting layers can contribute significantly to the inverse localization length, i.e. reduce dramatically the total transmission at typical realizations.

In calculating the inverse localization length for $S$-wave we follow the general procedure presented in Sec. 2, and start from Eq. (4). Since both transparent ($tr$) and reflecting ($ref$) layers are present, it is advantageous to separate the sum in Eq. (4) into two, each of them related to a particular type of layers. In doing this it is worthwhile to couple each reflecting layer with its left-side transparent neighbor. The number of such pairs is equal to the number of reflecting slabs, $N_{ref}$, while the amount of the remaining (uncoupled) transparent slabs is $N_{tr} - N_{ref}$ ($N_{tr}$ is the number of transparent layers). Taking this into account we can rewrite Eq. (4) as

$$l_{loc}^{-1} = -\lim_{N \to \infty} L^{-1} \left( \sum_{j=1}^{N_{tr} - N_{ref}} \ln \frac{A_{j+1}}{A_j} + \sum_{j=1}^{2N_{ref}} \ln \frac{A_{j+1}}{A_{j-1}} \right) \ . \tag{31}$$

Under assumption that the mean thicknesses of reflected and transparent layers are the same, $s$, the expression Eq. (5) for the inverse localization length takes the form

$$l_{loc}^{-1} = -\left( \frac{N_{tr} - N_{ref}}{Ns} \left\langle \ln \frac{A_{j+1}}{A_j} \right\rangle_{tr} + \frac{N_{ref}}{Ns} \left\langle \ln \frac{A_{j+1}}{A_{j-1}} \right\rangle_{ref} \right) \equiv \frac{N_{tr} - N_{ref}}{N} l_{loc_{tr}}^{-1} + \frac{2N_{ref}}{N} l_{loc_{ref}}^{-1} \ . \tag{32}$$

Here $l_{loc_{tr}}^{-1}$ stands for the inverse localization length in the medium that consist of transparent layers only, while

$$l_{loc_{ref}}^{-1} = (2s)^{-1} \left\langle \ln \frac{A_{j-1}}{A_{j+1}} \right\rangle_{ref} \tag{33}$$

denotes the inverse localization length in the medium built of alternating transparent and reflected layers. Given the distribution function of the refractive index, $F_n(n)$, the numbers of layers can be calculated as

$$N_{tr} = N \int_{n' > n_c} F_n(n') dn' \ , \quad N_{ref} = N \int_{n' < n_c} F_n(n') dn' \ . \tag{34}$$

When $N_{ref} = 0$ or $l_{loc_{ref}}^{-1} = l_{loc_{tr}}^{-1}$, Eq. (32) turns, as it must, into Eq. (5).

To find the ratios of the amplitudes involved in Eq. (32) we note that the electric field, $E_j$, inside $j$th reflected layer is a superposition of two evanescent modes:

$$E_j = A_j \exp(-\gamma_j z) + B_j \exp(\gamma_j z) \ , \tag{35}$$

with

$$\gamma_j = -ik_{z_j} = \sqrt{k_x^2 - n_j^2 k_0^2} = k_0 \sqrt{n_c^2 - n_j^2} \ . \tag{36}$$

Electric field in the adjacent $(j-1)$th and $(j+1)$th transparent layers is given by Eq. (14). The explicit form of general connections (6) follows from the conditions of continuity of the fields and their derivatives at the boundaries

$$
\begin{aligned}
& A_{j-1} \cos \varphi_{j-1} = A_j + B_j \exp(-\gamma_j s_j), \\
& k_{z_{j-1}} A_{j-1} \sin \varphi_{j-1} = \gamma_j A_j - \gamma_j B_j \exp(-\gamma_j s_j), \\
& A_j \exp(-\gamma_j s_j) + B_j = A_{j+1} \cos(\varphi_{j+1} - \Delta_{j+1}), \\
& \gamma_j A_j \exp(-\gamma_j s_j) - \gamma_j B_j = k_{z_{j+1}} A_{j+1} \sin(\varphi_{j+1} - \Delta_{j+1}).
\end{aligned} \tag{37}
$$

From Eqs. (37) one can derive:

$$
\begin{aligned}
\frac{A_{j-1}^2}{A_{j+1}^2} &= \frac{1}{4\Gamma_j^2} \left( \cos \tilde{\varphi}_{j+1} - \frac{k_{z_{j+1}}}{\gamma_j} \sin \tilde{\varphi}_{j+1} \right)^2 \left( 1 + \frac{\gamma_j^2}{k_{z_{j-1}}^2} \right) \\
&+ \frac{1}{2} \left( \cos^2 \tilde{\varphi}_{j+1} - \frac{k_{z_{j+1}}^2}{\gamma_j^2} \sin^2 \tilde{\varphi}_{j+1} \right) \left( 1 - \frac{\gamma_j^2}{k_{z_{j-1}}^2} \right) + \frac{\Gamma_j^2}{4} \left( \cos \tilde{\varphi}_{j+1} + \frac{k_{z_{j+1}}}{\gamma_j} \sin \tilde{\varphi}_{j+1} \right)^2 \left( 1 + \frac{\gamma_j^2}{k_{z_{j-1}}^2} \right),
\end{aligned} \tag{38}
$$

where $\Gamma_j = \exp(-\gamma_j s_j) \ll 1$, and $\tilde{\varphi}_{j+1} = \varphi_{j+1} - \Delta_{j+1}$.

Thus, to calculate $l_{loc_{ref}}^{-1}$ one has to perform the averaging in Eq. (33) with $A_{j-1}/A_{j+1}$ substituted from Eq. (38). In accordance with Eq. (8) in the short-wave limit we first average over uniformly distributed random phases $\tilde{\varphi}_{j+1}$, that in the first approximation in small parameter $\Gamma_j \ll 1$ after rather cumbersome calculations yields

$$\frac{1}{2\pi} \int_0^{2\pi} \ln \frac{A_{j+1}}{A_{j-1}} d\tilde{\varphi} = \ln \Gamma_j [1 + O(\Gamma_j^2)] \approx \gamma_j s_j \ . \tag{39}$$

Eq. (39) shows that change of the amplitude caused by the tunneling through a reflected layer is determined mainly by the attenuation rate $\gamma$ inside the layer, and is practically independent on the parameters of the adjacent transparent layers.

Finally, taking into account Eqs. (8), (33), (36), and (39) for the contribution of the reflecting layers to the inverse localization length we obtain

$$l_{loc_{ref}}^{-1} = (2s)^{-1} \int ds' F_s(s') \int_{n' < n_c} dn' F_n(n') \, s' \gamma(n') = \frac{k_0}{2} \int_{n' < n_c} dn' F_n(n') \sqrt{n_c^2 - n'^2} \quad . \tag{40}$$

Evidently, this contribution does not depend on the wave polarization. Therefore, the resulting inverse localization length for $S$-polarized or $P$-polarized waves can be obtained from Eq. (32) with $l_{loc_{tr}}^{-1}$ substituted by Eqs. (20) or (21) respectively.

When $N_{ref} / N_{tr} \ll 1$ (in the opposite limit the effect of transparent layers is negligible and the exponential decay of the transmission coefficient has nothing to do with localization), Eq. (32) with account made for Eqs. (34), (40) transforms to

$$l_{loc}^{-1} \approx l_{loc_{tr}}^{-1} + \delta \quad , \quad \delta = \frac{2N_{ref}}{N} l_{loc_{ref}}^{-1} = k_0 \int_{n' < n_c} F_n(n') dn' \int_{n'' < n_c} F_n(n'') \sqrt{n_c^2 - n''^2} \, dn'' \quad . \tag{41}$$

Easy to see, the ratio $\delta / l_{loc_{ref}} \propto l_{loc_{ref}} / l_{loc_{tr}}$ is proportional to $k_0 s \gg 1$, which means that the influence of the reflecting layers can be significant even when their number is small.

To demonstrate characteristic physical features of the localization in the presence of reflecting layers we consider a medium consisted of statistically independent random layers with rectangle distribution function of the refractive index, $F_n$:

$$F_n(x) = \begin{bmatrix} (n_{max} - n_{min})^{-1} \, , \; n_{min} < x < n_{max} \\ 0 \, , \; x < n_{min} \text{ or } x > n_{max} \end{bmatrix} \quad . \tag{42}$$

Reflecting layers exist if $n_{min} < n_c$. When $1 - n_{min} / n_c \ll 1$, the relative numbers of reflecting and transparent slabs are

$$\frac{N_{ref}}{N} = \frac{n_c - n_{min}}{n_{max} - n_{min}} \ll 1 \, , \quad \frac{N_{tr}}{N} = \frac{n_{max} - n_c}{n_{max} - n_{min}} \approx 1 \quad . \tag{43}$$

In this case after substitution of Eq. (42) in Eq. (41) one obtains

$$\delta = \frac{k_0 (n_c - n_{min})}{2(n_{max} - n_{min})^2} \int_{n_{min}}^{n_c} \sqrt{n_c^2 - n'^2} \, dn' = \frac{k_0 n_c^2 (n_c - n_{min})}{2(n_{max} - n_{min})^2} \arcsin\left( \frac{n'}{n_c} \right) \Big|_{n_{min}}^{n_c}$$

$$\approx \frac{k_0 n_c^3}{\sqrt{2}(n_{max} - n_{min})^2} \left( 1 - \frac{n_{min}}{n_c} \right)^{\frac{3}{2}} \quad . \tag{44}$$

The contribution to the inverse localization length from transparent layers, $l_{loc_{tr}}^{-1}$, is given by Eqs. (20) and (21). If, for example, the values of $n_{max}$ and $n_{max} - n_{min}$ are of the order of one, $n_{max} \sim n_{max} - n_{min} \sim 1$, it is easy to show that

$$l_{loc_{tr}}^{-1} \sim s^{-1} \quad . \tag{45}$$

From comparison Eq. (45) with the Eq. (44) the following conditions can be obtained for the contribution of the reflecting layers to be of the same order as that of the transparent ones:

$$n_{min} \sim n_c - \frac{1}{n_c (k_0 s)^{2/3}} \quad , \tag{46}$$

when the angle of propagation $\vartheta_0 \sim n_c$ is fixed, and

$$\vartheta_0 \sim \frac{n_{min}}{2} + \sqrt{\frac{n_{min}^2}{4} + \frac{1}{(k_0 s)^{2/3}}} \quad , \tag{47}$$

when $n_{min}$ is fixed. Note, that angle $\vartheta_0$ in Eq. (47) is always larger than $(k_0 s)^{-1/3}$. Hence, when $\vartheta_0 < (k_0 s)^{-1/3} \ll 1$ the influence of the reflecting layers is small independently on $n_{min}$.

Fig. 1 shows the typical dependences $l_{loc_{S,P}}^{-1}(\vartheta_0)$ calculated numerically for the rectangle distribution function Eq. (42).

## 5. CONCLUSION

Transmission of a plane monochromatic transverse wave obliquely incident on a randomly layered medium has been studied, and the inverse localization length determined as $l_{loc}^{-1} = -\langle \ln T / 2L \rangle$ has been calculated in the high-frequency (strong disorder) limit. The method of calculation takes advantage of the fact that in the localization regime the energy flux at typical realizations is zero (with exponential accuracy), and therefore the field inside each layer can be considered as a standing wave. The assumption is also used that the phases at the interfaces of layers are uniformly distributed in the interval $[0, 2\pi)$. The approach is rather general, and simplifies calculations significantly as compared to the conventional transfer matrix approach. With this method we have shown that the inverse localization length of $S$-waves increases with the angle of propagation, and is always larger than the localization length of $P$-waves, which, in contrast, goes down as the angle of incidence grows, and reaches a minimum at some angle (stochastic analog of Brewster effect). The effect is most pronounced in the medium consisted of alternating layers of two dielectrics with random widths. In this case $l_{loc} \to \infty$ at the Brewster angle, i.e. the localization is absent for $P$-waves. If in a random sample there are layers with the refractive index small enough, $n < \sin \vartheta_0$, strong internal reflection from these layers can reduce significantly the total transmission. This reduction is described by an additive term in the inverse localization length, which depends on the number of layers that are reflecting at the given angle of incidence, and on the statistics of their parameters (width and dielectric constants). The conditions for this term to be comparable with the contribution from transparent layers have been discussed. The revealed dependence of the localization length on the angle of propagation could play a vital part in formation of the field of a source radiating in different directions, and enhance significantly the waveguiding effect in randomly layered media [12].

## ACKNOWLEGMENT

The work was partially supported by INTAS (grant 03-55-1921) and Israeli Science Foundation (grant 328/02).

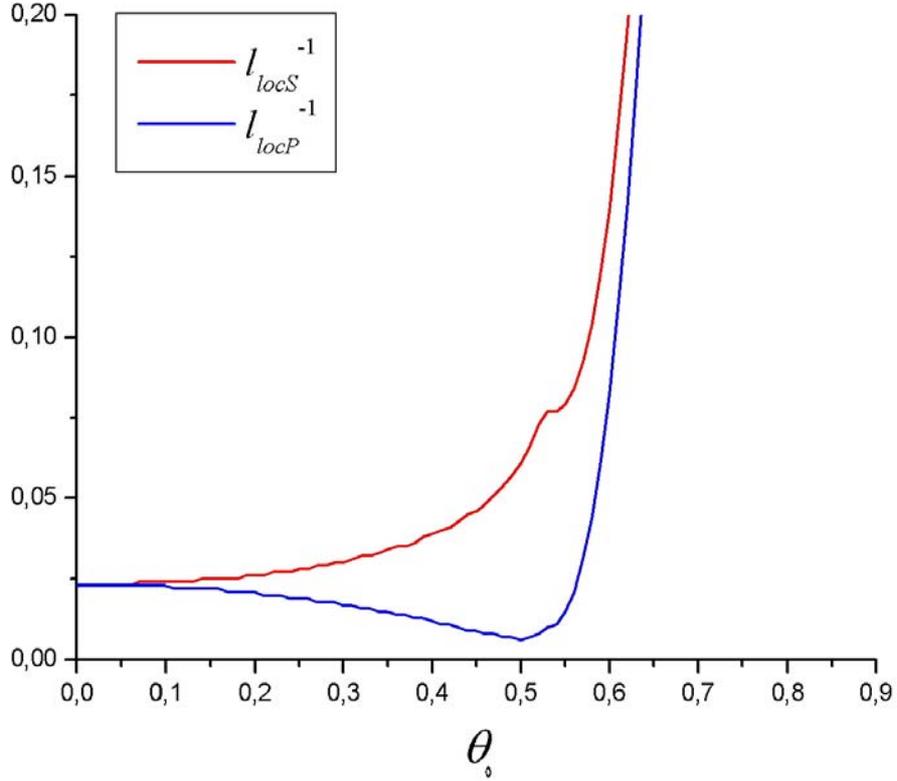

**Fig. 1.** Dependences $l_{loc_S}^{-1}$ and $l_{loc_P}^{-1}$ on the angle of incidence, $\vartheta_0$. The curves are obtained by numerical calculations of Eqs. (41), (20) and (21) with rectangle distribution function (42) and the following values of the parameters: $n_{\min} = 0.5$, $n_{\max} = 1.5$, $s = 1$, and $k_0 = 100$. At $\vartheta_0 = \tilde{\vartheta}_B \approx 0.5$ function $l_{loc_P}^{-1}(\vartheta_0)$ has minimum; the angle $\tilde{\vartheta}_B$ is an analogue of the Brewster angle. At $\vartheta_0 > \arcsin n_{\min} \approx 0.52$ both functions rapidly increase because of the influence of reflected layers, which increase in number with $\vartheta_0$ increasing.